\documentclass[preprint]{JASA}

\begin{document}

\title[A Scalable Approach to Estimating the Rank of High-Dimensional Data]{A Scalable Approach to Estimating the Rank of High-Dimensional Data}

\author{Wenlan Zang}
\email{wenlan.zang@yale.edu}
\author{Jen-hwa Chu}
\altaffiliation{Also at: Section of Pulmonary, Critical Care and Sleep Medicine (PCCSM), Yale University, New Haven, CT 06510, USA}
\author{Michael J. Kane}
\affiliation{Department of Biostatistics, Yale University, New Haven, CT 06510, USA}



\date{\today} 
\renewcommand{\linenumbers}{}
\begin{abstract}
A key challenge to performing effective analyses of high-dimensional data is finding a signal-rich, low-dimensional representation. For linear subspaces, this is generally performed by decomposing a design matrix (via eigenvalue or singular value decomposition) into orthogonal components, and then retaining those components with sufficient variations. This is equivalent to estimating the rank of the matrix and deciding which components to retain is generally carried out using heuristic or ad-hoc approaches such as plotting the decreasing sequence of the eigenvalues and looking for the ``elbow'' in the plot. While these approaches have been shown to be effective, a poorly calibrated or misjudged elbow location can result in an overabundance of noise or an under-abundance of signal in the low-dimensional representation, making subsequent modeling difficult. In this article, we propose a latent-space-construction procedure to estimate the rank of the {\em detectable} signal space of a matrix by retaining components whose variations are significantly greater than random matrices, of which eigenvalues follow a universal March\u{e}nko-Pastur (MP) distribution. 
\end{abstract}

\maketitle

\section{\label{sec:1} Introduction and Background}
High-dimensional data analysis, in which the number of variables or features in a data set is larger than the number of samples, is common in signal processing, computer vision, and genomic studies \citep{yang1996asymptotic, white2005spectral}. For these types of data sets, it is generally assumed that there is a low-dimensional and information-rich subspace within the data set \citep{hoff2007model}. Subsequent methods prescribe a variety of dimension-reduction techniques to find a linear vector subspace that provides a low-dimensional representation of high-dimensional data and retaining those components of the data carrying predictive (or otherwise significant) information. A standard construction for this class of problems considers decomposing an $n \times p$ matrix $X$ into a rank $K$ $(K \ll p)$ signal matrix $S$ plus a noise matrix $N$. The determination of $K$ is paramount for dimension-reduction problems to avoid over- or under-fitting subsequent models \citep{carreira1997review}. However, due to the challenge of dimension estimation and relatively few effective approaches proposed, many models take rank $K$ as an analyst-specified parameter based on ill-defined or ad-hoc methods. For example, a standard method of estimating rank $K$ is to retain components that explain 70\% of total variance or to plot the decreasing sequence of the matrix's eigenvalues (i.e., scree plot) and having an individual pick the point at which the rate of decrease itself drops significantly (i.e., ``elbow’’) \citep{yong2013beginner}. The eigenvalue index at which the elbow occurs is denoted as $K$. An alternative approach (jackstraw) implements a statistical test to identify features with a statistically significant association with any linear combination of principal components (PCs) of interest based on a constructed null distribution by randomly permuting the data set \citep{chung2015statistical}. Although jackstraw is widely applied in dimension estimation, it requires the number of PCs that capture ``systematic variation from latent variables" as an input for the algorithm, and the test results from jackstraw only provide a subset of suggested cutoffs from input PCs instead of a clear dimension-estimation cutoff \citep{chung2015statistical}. Furthermore, jackstraw is computationally intensive and time-consuming when applied to large data sets.

Several attempts have been made previously to solve the rank estimation problem. \citet{hoff2007model} proposed using the Bayesian rank-estimation procedure based on the signal-plus-Gaussian-noise model, providing prior and posterior distribution of rank $K$ and posterior estimation of right- and left- singular vectors of matrix $X$. Posterior inferences can be drawn from expectation $E_X[p(K|X)]$ and rank can be determined by the posterior mode of $p(K|X)$. The major weakness that prohibits application of Hoff’s method to large data sets (e.g., 1,000 nodes) is the computational complexity of the markov chain monte carlo (MCMC) algorithm \citep{hoff2007model}. In addition, in the markov chain, \citet{hoff2007model} sampled eigenvalues and the corresponding eigenvectors from the full condition distribution instead of the marginal distribution. As a result, the probabilities of sampling nonzero eigenvalues do not mix well across different ranks of the mean matrix, as eigenvalues and the corresponding eigenvectors are dependent \citep{hoff2007model}. Recently, \citet{ladle} introduced a ladle estimator that combined the decreasing sequence of eigenvalues with the increasing pattern of variability of eigenvector directions. They observed that bootstrap variability of the eigenvectors decrease as the eigenvalues get further apart and increase as the eigenvalues get closer. \citet{ladle} proposed the ladle estimator based on this observation. They demonstrated the consistency of the ladle estimator and its superior performance in solving various dimension-reduction problems in a low-dimensional setting. While Hoff's method and ladle have been shown to be effective in several settings, if the focus is on large or high-dimensional data, then these two methods may not be the best choice for rank estimation.

We propose a procedure to estimate the rank of the detectable signal space of a high-dimensional matrix using the signal-plus-noise model under the assumption that the eigenvalues of the noise matrix follow the universal behavior described in \citet{aparicio2018quasi}. Under this construction, the expected spectrum of the random matrix $N$ follows a universal MP distribution, which can be compared to the eigenvalues of $X$ \citep{Mar_enko_1967}. In this context, components carrying {\em detectable} signal can be characterized as those whose eigenvalues are significantly larger than expected under the MP distribution. This construction provides a method for estimating the dimension of a data set and extracting those components of the data that may not be attributable to randomness. The major contribution of this chapter is to present this novel and practical method, which results in improved accuracy, robustness, and computational efficiency, especially for high-dimensional data.

The paper is organized as follows: section~\ref{sec:2} describes the framework setup and provides a distributional signal-to-ratio bound. Section~\ref{sec:3} illustrates a procedure of estimating detectable signal dimension and demonstrates its performance in relation to other standard approaches, including the ladle and Hoff’s estimators. Section~\ref{sec:4} presents an application of dimension-estimation method in scRNA-seq data analysis. Section~\ref{sec:5} concludes with a discussion.

\section{\label{sec:2} A bound on the detectable signal dimension }
\subsection{\label{subsec:2:1} Signal-Plus-Noise Model}
Let $X_{n \times p}$ be a matrix with mean-centered columns consisting of a low-rank matrix $S$ plus a random matrix $N$. Without loss of generality, we assume $p > n$ for high-dimensional data. Following the construction of \citet{livan2018introduction}, assume matrix $N$ is formed by $n \times p$ samples from $p$ random variables $[x_1, ..., x_p]$ drawn from an unknown joint distribution $p(x)$. Let $N_{ij}$ be independent and identically distributed (i.i.d.) entries with $\E(N_{ij}) = 0$ and $\E({N^2}_{ij}) = \sigma^2$ \citep{gotze2004rate}. The 
scaled inner product of $X$ can be written as
\begin{align}
    \frac{1}{p} XX^T &= \frac{1}{p}(S + N)(S + N)^T \\
         &= \frac{1}{p}SS^T + \frac{1}{p}NN^T + \frac{1}{p}SN^T + \frac{1}{p}NS^T \,. \notag
\end{align}
We can specify finding the rank of $X$ in terms of eigenvalue decomposition as follows:
\begin{equation}
XX^T = U_X \Lambda_X U_X^T\,,
\end{equation}
where $u_{X_i}$ denotes the $i$th column of the eigenvector $U_X$ and $\Lambda_X$ is a diagonal matrix with diagonal entries $\{\lambda_{X_1}, ..., \lambda_{X_n}\}$ as the non-negative eigenvalues sorted in a non-increasing order. When $n \geq p$, the $\Lambda_X = diag\{ \lambda_{X_1}, ..., \lambda_{X_p},0,...,0 \}$, where $n -p$ eigenvalues equals zero. Thus, we can rewrite the scaled eigenvalues of matrix $XX^T$ in terms of the following approximation:
\begin{align} \label{dimsimp}
    \frac{1}{p}\Lambda_X(X) &= \frac{1}{p} U_X^T \left( XX^T \right) U_X \\ \notag
    &= \frac{1}{p}U_X^T \left( SS^T + NN^T + SN^T  + NS^T \right) U_X \\ \notag
    &\approx \frac{1}{p} U_X^T \left( U_S \Lambda_S(S) U_S^T + U_N \Lambda_N(N) U_N^T \right)
    U_X \\ \notag
    &\approx \frac{1}{p}\Lambda_S(S) + \frac{1}{p}\Lambda_N(N) \,. \notag
\end{align}
There are two challenges in the simplification provided in Equation \ref{dimsimp}. First, adding $N$ to $S$ can result in the underestimation of the elements in $\Lambda_S(S)$ using $\Lambda_X(S)$. Second, the eigen vectors of $S$ and $N$ are not generally well-aligned ($U_N^T U_S \not\approx I$). As a result, we provide rigorous justification below and in Appendix \ref{appendix:a}. Given $S$ and $N$ are independent, we have $\frac{1}{p}SN^T$ and $\frac{1}{p}NS^T$ converge to 0 at a rate of $\frac{\Tr (S)}{p}$ when $p \to \infty$ (see Appendix \ref{appendix:a}). $\Lambda_S(S)$ is a rank $K$ diagonal matrix with non-zero entries corresponding to the  rank of the signal space. When the signal's magnitude is large compared to the noise, the corresponding vectors in $U_X$ are aligned with $U_S$ in the first $K$ dimensions. $U_X$ can therefore be regarded as as a perturbation on $SS^T$, and $U_X^TU_S$ will be approximately identical in the $K$ leading eigenvectors. The remaining $(n-K)$ dimensions have eigenvalues close to zero. At the same time, $W_p = \frac{1}{p}NN^T$ is a random matrix converging to $I$ (see Appendix \ref{appendix:b}) and $U_X^T N^TN U_X$ can be seen as a rotation on the noise matrix, implying that $U_X^T U_N \approx I$.

When $n \geq p$, we calculate the scaled inner product of $W_n = \frac{1}{n} X^TX$. Similarly, the scaled eigenvalues of $X^TX$ can be rewritten as 
\begin{align}
\frac{1}{n}\Lambda_X(X) &\approx \frac{1}{n}\Lambda_S(S) + \frac{1}{n}\Lambda_N(N)\,.
\end{align}
The top $p$ eigenvalues of $X^TX$ and $XX^T$ are the same up to a constant $n$ or $p$ and the rest $n-p$ eigenvalues are equal to zero for $n \geq p$. We only focus on the consistent eigenvalues up to the minimum of $n$ or $p$. The eigenvalues in the null space are not considered in this paper. Similarly, we can demonstrate that $\frac{1}{n}S^TN = \frac{1}{n}N^TS \approx 0$ and that $\frac{1}{n}N^TN$ converges to $I$.

\subsection{\label{subsec:2:2} Empirical Spectral Distribution}
By providing an estimate of $\Lambda_N(N)$, we estimate the amount of detectable signal that can be distinguished from noise in the region where $\Lambda_X(X) - \Lambda_N(N) \geq 0$. We denote $W = NN^T$ and its $n$ non-negative eigenvalues as $[\lambda_1,\lambda_2,...,\lambda_n]$. The joint density of eigenvalues for the Wishart matrices is as follows:
\begin{align}
\label{jpdf}
p(\lambda_1,\lambda_2,...,\lambda_n)&={Z^{(L)}_{n,\beta}}e^{- \beta \sigma\sum_{i=1}^n{\lambda_i}}\prod_{i=1}^{n}{\lambda_i}^{\frac{\beta}{2}[(p-n+1)-\frac{2}{\beta}]}\prod_{j<k}{|\lambda_j-\lambda_k|}^{\beta}\\
&={Z^{(L)}_{n,\beta=1}}e^{-\frac{1}{2}\sum_{i=1}^n{\lambda_i}}\prod_{i=1}^{n}{\lambda_i}^{\frac{1}{2}(p-n-1)}\prod_{j<k}{|\lambda_j-\lambda_k|} \,, \notag
\end{align}
\noindent where $\beta=1$ for real entries and $\sigma=\frac{1}{2}$ for standard normal distribution \citep{gotze2004rate, livan2018introduction}. When $n \geq p$, we define it as Anti-Wishart ensemble $\tilde{W}$, where we have $n - p$ eigenvalues equal zero. The joint density function of the Anti-Wishart ensemble is similar to Equation \ref{jpdf}, with part of the matrix elements being random and the remaining elements non-random. The determination of these non-random elements are related to the first $p$ rows of $\tilde{W}$ \citep{livan2018introduction}. The normalization factor can be calculated by Selberg's integral as
\begin{align}
{Z^{(L)}_{n,\beta =1}}&=(\frac{\beta}{2})^{\gamma}\prod_{j=1}^{n} {\frac{\Gamma{\big(\frac{\beta}{2}+1\big)}}{\Gamma{\big(\frac{\beta}{2}j+1\big)}\Gamma{\big(\frac{\beta}{2}(j-1)+\alpha+1\big)}}}\\
&=(\frac{1}{2})^{\frac{np}{2}}\prod_{j=1}^{n} {\frac{\Gamma{\big(\frac{3}{2}\big)}}{\Gamma{\big(\frac{j}{2}+1\big)}\Gamma{\big(\frac{1}{2}(p-n+j)\big)}}}\,, \notag
\end{align}
where $\gamma=n[\alpha+\frac{\beta}{2}(n-1)+1]$ and $\alpha=\frac{\beta}{2}(p-n+1)-1$ \citep{livan2018introduction,forrester2008importance, kumar2019recursion, dumitriu2005eigenvalues}.
\noindent The empirical spectral distribution of $W_p$ is defined as 
\begin{equation}
    F_n(y) = \frac{1}{n}\sum_{k=1}^n I_{\{\lambda_k \leq y\}}\,.
\end{equation}
\noindent As proven in \citet{Mar_enko_1967, gotze2004rate, livan2018introduction}, the expected spectral distribution $EF_n(y)$ and $F_n(y)$ converge to the MP distribution in probability with density form:
\begin{equation}
f_c(y)=\frac{1}{2\pi\sigma^2 cy}\sqrt{(y-c_{-})(c_{+}-y))} I_{[c_{-},c_{+}]}(y) + I_{[1,\infty)}(c)(1-c^{-1})\delta(y)\,,
\end{equation}
\noindent where rectangularity ratio $lim_{n,p \to \infty} (n/p) \to c \in (0, \infty)$. Note that ratio of $n/p$ approximates $c$ and is not fixed. Denote $\delta(y)$ as the Dirac delta function and $c_{\pm}=\sigma^2(1 \pm \sqrt{c})^2$ \citep{livan2018introduction, gotze2004rate, vivo2008wishart}. When $p > n$, we have the rectangularity ratio $c < 1$ and the second term in the density form equals zero according to the index function. When $n \geq p$, we have the rectangularity ratio $c \geq 1$ and the second term of the density function does not equal zero only when the eigenvalues equal zero. Note that eigenvalues in the null space are not discussed, as we assume that the signal space is smaller than the minimum of $n$ and $p$. As a result, the Dirac delta function equals zero for both high- and low-dimensional settings and share the same density form, which is as
\begin{equation}
f_c(y)=\frac{1}{2\pi\sigma^2 cy}\sqrt{(y-c_{-})(c_{+}-y))} I_{[c_{-},c_{+}]}(y)\,.
\end{equation}

\subsection{\label{subsec:2:3} Signal-to-Noise Ratio Bound}
The empirical spectral distribution provides an estimate of the eigenvalues $\Lambda_N(N)$ for the $W_p$ matrix in high- and low-dimensional cases. We focus on the high-dimensional case where $p > n$ in the in-text derivation and provide a distributional upper bound on the maximum eigenvalue of $\Lambda_N(N)$ to determine the magnitude of the signal that can be distinguished from noise. The derivation of the signal-to-noise ratio (SNR) bound for low-dimensional cases can be derived similarly.

The proof of the SNR bound can be summarized as follows: (i) provide the distribution of the diagonal elements of the $W_p$ matrix. (ii) Provide the distribution of the off-diagonal elements of the $W_p$ matrix and show that the off-diagonal elements converge to zero. (iii) Set the bound for the eigenvalues of the $W_p$ matrix by the Gershgorin circle theorem \citep{gerschgorin1931bounding}. Intuitively, the eigenvalue of the $j$th column of the $W_p$ square matrix lies within a circle of the $j$th diagonal elements at a radius of the sum of the absolute values of off-diagonal elements of the $j$th column \citep{gerschgorin1931bounding}. (iv) Further set the bound for the maximum eigenvalue of the $W_p$ matrix by the order statistics \citep{frechet1927loi}.

Assuming that the entries of the noise matrix $N$ are i.i.d. random variables drawn from an unknown joint distribution (without normality assumption) and $p$ is sufficiently large, we first note that by the central limit theorem, the diagonal and off-diagonal entries of $W_p$ matrix converge in distribution to normal distributions respectively, as $n$ and $m$ goes to infinity, where $m = n(n-1)/2$ is the number of elements in the upper triangular matrix.

\begin{theorem}
\label{theorem: diagnal distribution}
Let $N \in \mathcal{R}^{n \times p}$ with i.i.d. elements having mean 0, variance $\sigma^2$, and the fourth moment $\gamma^4 < \infty$. The diagonal entries of the $W_p$ matrix $\xi_j$ $(j = 1, ..., n)$ converge in distribution to $\mathcal{N}(\sigma^2, \frac{\gamma^4 - \sigma^4}{p})$ when $p$ is sufficiently large to have the central limit effect.
\begin{proof}\renewcommand{\qedsymbol}{}
see Appendix \ref{appendix:c}.
\end{proof}
\end{theorem}

\begin{theorem}
\label{theorem: off-diagnal distribution}
Let $N \in \mathcal{R}^{n \times p}$ with i.i.d. elements having mean 0, variance $\sigma^2$, and the fourth moment $\gamma^4 < \infty$. The off-diagonal entries of the $W_p$ matrix $\varepsilon_{ik}$ ($i,k = 1, ..., n$ and $i<k$) converge in distribution to $\mathcal{N}(0, \frac{\gamma^4 + \sigma^4}{4p})$ when $p$ is sufficiently large to have the central limit effect.
\begin{proof}\renewcommand{\qedsymbol}{}
see Appendix \ref{appendix:d}.
\end{proof}
\end{theorem}

Consider the special case where $N$ follows a standard normal distribution, the variance of the main diagonal elements is $\frac{\gamma^4 - \sigma^4}{p} = \frac{3\sigma^4 - \sigma^4}{p} = \frac{2}{p}$, or similarly by the main diagonal of variance for $W_p$ as $\frac{1}{p}(n+1) = \frac{2}{p}$. Thus, the distribution of the main diagonal elements is $\mathcal{N} (1, \frac{2}{p})$, the justification for which is rigorously provided in Appendices \ref{appendix:b} and \ref{appendix:c}. Similarly, the distribution of the off-diagonal elements is $\mathcal{N} (0, \frac{1}{p})$, which is rigorously justified in Appendix \ref{appendix:d}. As proved in Appendix \ref{appendix:b}, the first moment of the $W_p$ matrix for the normalized noise matrix is the identity matrix, with the off-diagonal entries converging to zero at a rate of $\frac{1}{\sqrt{p}}$ when $p$ is sufficiently large.

Theorems \ref{theorem: diagnal distribution} and \ref{theorem: off-diagnal distribution} provide the distribution of the diagonal and off-diagonal entries of the $W_p$ matrix. We use the results of order statistics to estimate an upper bound of the maximum diagonal element and the maximum radius of the Gershgorin circle of $W_p$ matrix. A similar upper bound can be estimated for the diagonal and the maximum radius of the Gershgorin circle for $W_n$ matrix when $n \geq p$.

\begin{theorem}
\label{theorem: diagnal bound}
Let $\xi_j$ be the main diagonal elements of $W_p$, where $j = 1, ..., n$. The Fisher–Tippett–Gnedenko theorem shows that for the maximum diagonal element, $\xi_{(p)} = max(\xi_j)$. We have
\begin{equation}
    \E \{ \xi_{(n)} > \alpha_1 \} = e^{-e^{-{\sqrt{p}(\alpha_1 - \sigma^2)}/{\sqrt{\gamma^4- \sigma^4}}}}\,.
\end{equation}
We can set the bound of $\xi_{(n)}$ with the inequality $e^{x} \leq e^{x^2} + x$ for $x \in \mathbb{R}$ such that
\begin{equation}
    \E \{ \xi_{(n)} > \alpha_1 \} \le e^{e^{-\frac{2\sqrt{p}(\alpha_1 - \sigma^2)}{\sqrt{\gamma^4- \sigma^4}}}} - e^{-\frac{\sqrt{p}(\alpha_1 - \sigma^2)}{\sqrt{\gamma^4- \sigma^4}}}\,.
\end{equation}
\begin{proof}\renewcommand{\qedsymbol}{}
    see Appendix \ref{appendix:e}.
\end{proof}
\end{theorem}

\begin{theorem}
\label{theorem: off-diagnal bound}
Let $\varepsilon_{ik}$ be the off-diagonal elements of $W_p$, which are normally distributed with mean 0 and variance $\frac{\gamma^4 + \sigma^4}{4p}$, where $i,k = 1, ..., n$ and $i<k$. Let $R_k = \sum_{i \neq k}|\varepsilon_{ik}|$ be the sum of the absolute values of the off-diagonal elements of the $k$th column of $W_p$ matrix with expectation $\mu_{GEV} = \sqrt{\frac{(n-1)^2(\gamma^4 + \sigma^4)}{2p\pi}}$ and variance $\sigma^2_{GEV}= \frac{(n-1)(\pi - 2)(\gamma^4 + \sigma^4)}{4p\pi}$. The Fisher–Tippett–Gnedenko theorem shows that for the maximum entry, $R_{(n)} = max(R_k)$. We have 
\begin{equation}
\E \{ R_{(n)} > \alpha_2 \} = e^{-e^{-b}}\,,
\end{equation}
where $b = \frac{\alpha_2 - \mu_{GEV}}{\sigma_{GEV}}$.
We can set the bound of $R_{(n)}$ with the inequality $e^{x} \leq e^{x^2} + x$ for $x \in \mathbb{R}$ such that
\begin{equation}
    e^{-e^{-b}} \leq e^{e^{-2b}} -e^{-b}\,.
\end{equation}
The eigenvalue of the $k$th column of the $W_p$ matrix can be bound by the Gershgorin circle theorem. With the distributional upper bound of the $\xi_k$ and distributional upper bound of $R_k$, we have the upper bound of the largest eigenvalue of $W_p$ matrix, which lies within the Gershgorin discs $D(\xi_{(n)},R_{(n)})$ that are centered at $\xi_{(n)}$ with a radius $R_{(n)}$ of rate of convergence $\sqrt{\frac{n-1}{p}}$ for $p > n$.
\begin{proof}\renewcommand{\qedsymbol}{}
    see Appendix \ref{appendix:f}.
\end{proof}
\end{theorem}

\begin{corollary}
\label{theorem: normal diagnal bound}
Let $N \in \mathcal{R}^{n \times p}$ with i.i.d. elements following a standard normal distribution with mean 0 and variance 1. The diagonal entries of the $W_p$ matrix $\xi_j$ $(j = 1, ..., n)$ converge in distribution to $\mathcal{N}(1, \frac{2}{p})$. The Fisher–Tippett–Gnedenko theorem shows that for the maximum diagonal element, $\xi_{(p)} = max(\xi_j)$. We have
\begin{equation}
    \E \{ \xi_{(n)} > \alpha_1 \} = e^{-e^{-\frac{\sqrt{2p}(\alpha_1 - 1)}{2}}}\,.
\end{equation}
we can set the bound of $\xi_{(n)}$ by 
\begin{equation}
    \E \{ \xi_{(n)} > \alpha_1 \} \le e^{e^{-\sqrt{2p}(\alpha_1 - 1)}} - e^{-\frac{\sqrt{2p}(\alpha_1 -1)}{2}}\,.
\end{equation}
The off-diagonal entries of the $W_p$ matrix $\varepsilon_{ik}$ ($i,k = 1, ..., n$ and $i<k$) converge in distribution to $\mathcal{N}(0, \frac{1}{p})$. Let $R_k = \sum_{i \neq k}|\varepsilon_{ik}|$ be the sum of the absolute values of the off-diagonal elements of the $k$th column of the $W_p$ matrix with expectation $\mu_{GEV} = \sqrt{\frac{2(n-1)^2}{p\pi}}$ and variance $\sigma^2_{GEV}= \frac{(n-1)(\pi - 2)}{p\pi}$. The Fisher–Tippett–Gnedenko theorem shows that for the maximum entry, $R_{(n)} = max(R_k)$. We have 
\begin{equation}
\E \{ R_{(n)} > \alpha_2 \} = e^{-e^{-b}}\,,
\end{equation}
where $b = \frac{\alpha_2 - \mu_{GEV}}{\sigma_{GEV}}$.
We can set the bound of $R_{(n)}$  with the inequality $e^{x} \leq e^{x^2} + x$ for $x \in \mathbb{R}$ such that
\begin{equation}
    e^{-e^{-b}} \leq e^{e^{-2b}} -e^{-b}\,.
\end{equation}
Thus, we can set the bound of the largest eigenvalue of $W_p$ by $D(\xi_{(n)},R_{(n)})$.
\end{corollary}

\section{\label{sec:3} Estimating the detectable signal dimension in practice}
\subsection{\label{subsec:3:1} Algorithm}
Equation \ref{dimsimp} shows that the rank of the signal can be estimated by first estimating $\Lambda_N(N)$ and then finding the zero crossing of the difference $\Lambda_X(X) - \Lambda_N(N)$ corresponding to where any signal in the signal space cannot be distinguished from noise (see Figure 1).

\begin{figure}[ht]
\includegraphics[width=\reprintcolumnwidth]{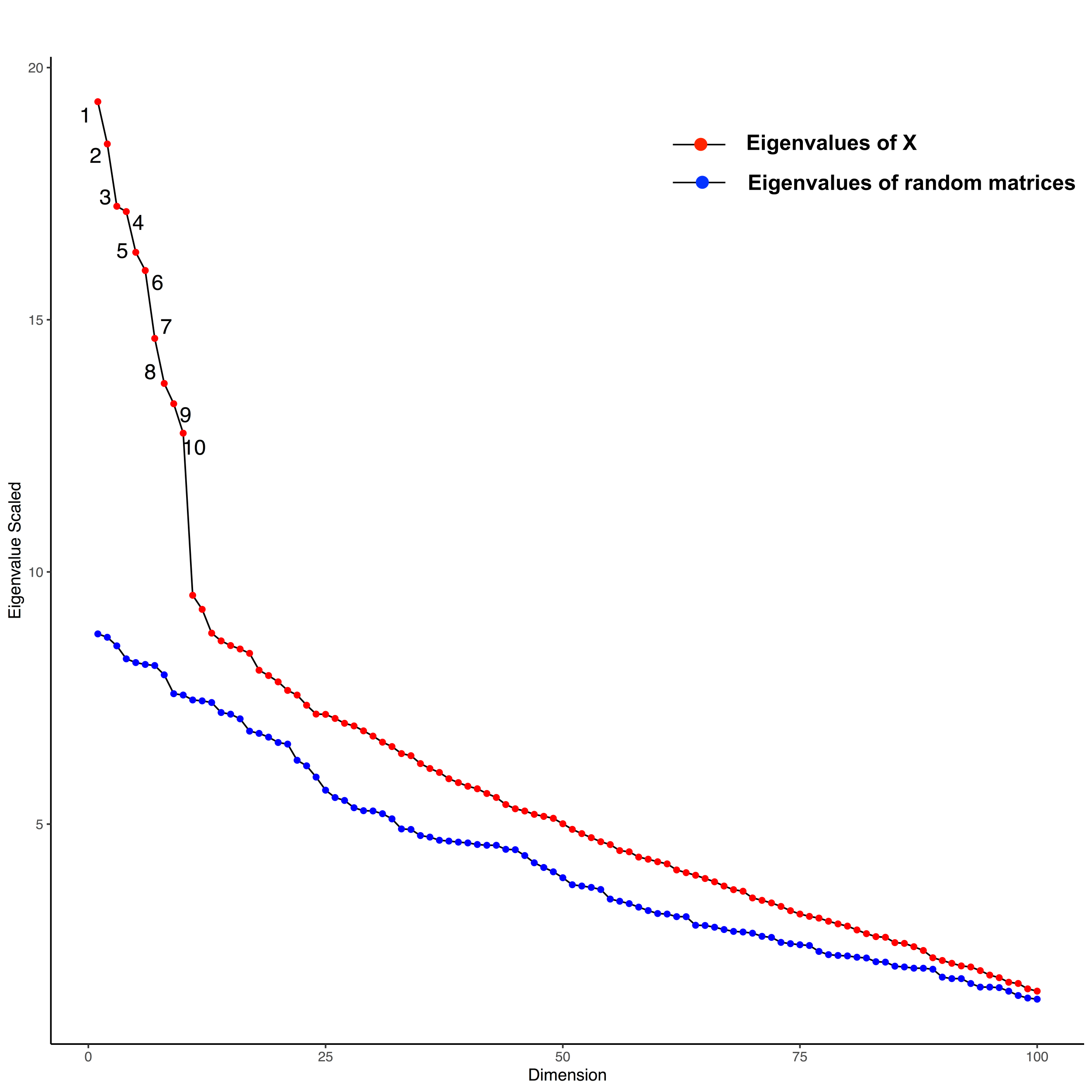}
\caption{\label{fig:FIG1}{Scaled eigenvalues of $X_{100 \times 500}$ with true rank 10 (red dotted line) compare to random samples from the MP distribution (blue dotted line).}}
\end{figure}

As a more concrete example, consider the case of plotting scaled eigenvalues of $X$ matrix with  100 samples, 500 features, and a true rank equals to 10 as well as random samples from the MP distribution in Figure 1. The red line plots the scaled eigenvalues of the matrix while the blue line plots the scaled eigenvalues of a similarly distributed noise-only matrix. The goal is to extract all the signal components up to $K$ whose eigenvalues are significantly larger than the expected eigenvalues from the MP distribution. The dimension of the detectable signal is the first eigenvalue index where the corresponding eigenvalue is close to the largest eigenvalue sampled from the MP distribution, meaning it cannot be distinguished from noise. The algorithm is implemented as follows:

\begin{algorithm}[H]
\caption{Algorithm for rank estimation}\label{alg:dimension}
\begin{algorithmic}[1]
\Procedure{Algorithm}{$X_{n \times p}$}
\State {Calculate eigenvalue decomposition for scaled matrix $X$ to dimension $n'$ ($n' \ll n$). Denote eigenvalues as $\Sigma_x$.}
\State {Draw random samples from the MP distribution with a corrected variance. Denote eigenvalues as $\Sigma_n$.}
\State {If $\exists \: \Sigma_n > \Sigma_x$, $K = 0$.}
\State {Calculate posterior probability of change points for deviation ($\Delta = \Sigma_x - \Sigma_n$) with a prior equals to a decreasing linear sequence from 0.9 to 0.}
\State Set $K$ to equal the largest dimension with the maximum posterior probability.
\EndProcedure
\end{algorithmic}
\end{algorithm}

\subsection{\label{subsec:3:2} Implementation}
An implementation of the dimension, ladle, and Hoff’s method as an open-source package for the R Programming environment can be found at \url{https://github.com/WenlanzZ/dimension}. We estimate the corrected variance of the MP distribution by the SNR bound to align the tails of the scaled eigenvalues of $X$ and noise. The test for deviation of scaled eigenvalues can be performed in several different ways. In practice, the univariate Bayesian change point (BCP) analysis, proposed by \citet{barry1993bayesian} has been found to be effective and is implemented in the R package bcp \citep{erdman2007bcp}. Frequently, there are several change points with their posterior probabilities approximating the maximum posterior probability. In order to guarantee the inclusion of all important signals, we propose a two-step procedure to estimate the dimension $K$ in practice. In the first step, we select candidates for $K$ by calculating the BCP on the posterior probability for a second time ({\em double posterior}). In the second step, among those candidates with the highest double posterior, we pick $K$ equal to the largest dimension where its posterior probability is larger than $\delta$ times the maximum posterior probability ($\delta \in \mathbb{R}:\delta \geq 0\text{ and }\delta \leq 1$ determine the range of approximation to the maximum posterior probability; by default set $\delta = 0.90$).

One limitation observed in simulation studies when implementing the dimension algorithm is related to R package bcp for calculating the posterior probabilities of the change points. The performance of R package bcp is not always stable at the edges of the sequence. Thus, in some extreme cases, a spike occurs at the edge after a long and flat pattern in the sequence of posterior probabilities. To address this problem, we adopted an alarm system that detects flat and spike patterns and trims off sequence before spikes or after a long and flat pattern.

\subsection{\label{subsec:3:3} Simulation}
We compared the performance of the proposed rank estimation procedure to the ladle and Hoff's method. Consider two signal-plus-noise matrices whose true signal rank $K = 3$. For a fair comparison, the first matrix setting is the same as in \citet{ladle} and the second matrix setting is similar to \citet{hoff2007model}:
\begin{subequations}
\begin{eqnarray}
X_1 \sim N(0, \Sigma_{X_1}), \Sigma_{X_1} = M + 0.54^2I_p, M = diag(2,1,1) + 0_{p,p}, \label{eqa}
\\
X_2 = S + \epsilon, S[\, \cdot \, , 1:K] \sim N(0, \Sigma), S[\, \cdot \,, (K+1):p] = 0, \Sigma=3 I_3. \label{eqc}
\end{eqnarray}
\end{subequations}
We simulated 1,000 independent samples from the $X$ matrix and performed rank estimation with different methods. We allowed each method to run for up to 24 hours for 1,000 simulations. The percentage of correct estimation and elapsed time in seconds are reported in Table 1. For example, in the first case of the $X_1$ matrix (sigma separated by commas) with signal rank equal to three and 100 samples and 10 features, dimension had 94.8\% of correct estimation with a mean absolute error of 0.056 and an average elapsed time of 0.040 seconds; ladle had 93\% of correct estimation with a mean absolute error of 0.396 and an average elapsed time of 0.034 seconds; Hoff's method had 78.8\% of correct estimation with a mean absolute error of 0.367 and an average elapsed time of 0.538 seconds. In the sixth case of the $X_2$ matrix (sigma as a single integer) with signal rank equal to three and 10 samples and 100 features, dimension had 28.3\% of correct estimation with a mean absolute error of 0.980 and an average elapsed time of 0.041 seconds; ladle had 0\% of correct estimation with a mean absolute error of 5.930 and an average elapsed time of 0.044 seconds; Hoff's method had 0\% of correct estimation with a mean absolute error of 7 and an average elapsed time of 1.094 seconds. In Table 1, when we fixed $p$ and let $n$ grow for the $n > p$ setting, the dimension method achieved the highest accuracy and performed stably for both the $X_1$ and $X_2$ matrices, while the ladle method seemed to work only with its own matrix setting. When we fixed $n$ and let $p$ grow for the $p > n$ setting, dimension exceled in both accuracy and computational efficiency. In terms of computational efficiency, the dimension method is more scalable to large data sets compared to the other methods. In extreme cases where noise overwhelms signal and the signal space is too small to be detectable, dimension underestimated the true rank by a mean absolute error of around 6. When signal is moderate in an extremely high-dimensional setting, dimension achieved an accuracy of 91.1\% with a mean absolute error of 0.207 and an average elapsed time of 3.767 seconds while the other two methods are computationally inapplicable to providing an estimation.

\begin{table}[ht]
\centering
\caption{\label{tab:table1} Simulation results comparing dimension estimation methods under different settings.}
\noindent\begin{tabular}{@{}*{13}{p{.076\textwidth}@{}}}
\hline\hline
\multicolumn{1}{l}{rank} & \multicolumn{3}{l}{unknown matrix} & \multicolumn{3}{l}{dimension} & \multicolumn{3}{l}{ladle} & \multicolumn{3}{l}{Hoff} \\
\cmidrule(lr){2-4}\cmidrule(lr){5-7}\cmidrule(lr){8-10}\cmidrule(lr){11-13}
k & n & p & sigma & acc & mae & time & acc & mae & time & acc & mae & time \\ \midrule
3 & 100 & 10 & 2, 1, 1 & 94.8 & 0.056 & 0.040 & 93 & 0.396 & 0.034 & 78.8 & 0.367 & 0.538\\[-1em]
3 & 1000 & 10 & 2, 1, 1 & 100 & 0 & 0.058 & 100 & 0 & 0.696 & 96.3 & 0.044 & 49.747\\[-1em]
3 & 10000 & 10 & 2, 1, 1 & 100 & 0 & 0.070 & 100 & 0 & 15.442 & - & - & \textgreater24h\\[-1em]
3 & 10 & 100 & 2, 1, 1 & 25.7 & 1.175 & 0.041 & 0 & 5.828 & 0.043 & 11.4 & 2.727 & 0.591\\[-1em]
3 & 10 & 1000 & 2, 1, 1 & 21.3 & 1.420 & 0.059 & 0 & 3.057 & 0.524 & - & - & \textgreater24h\\[-1em]
3 & 10 & 100 & 6 & 28.3 & 0.980 & 0.041 & 0 & 5.930 & 0.044 & 0 & 7 & 1.094 \\[-1em]
3 & 10 & 1000 & 6 & 18.2 & 1.696 & 0.062 & 0 & 3.123 & 0.493 & 0 & 7 & 116.187 \\[-1em]
3 & 10 & 10000 & 6 & 15.8 & 1.914 & 0.071 & 0 & 3 & 15.211 & - & - & \textgreater24h \\
\cline{1-13}
10 & 500 & 100 & 2, 2, 2 & 99.6 & 0.006 & 0.268 & 100 & 0 & 2.006 & 99.9 & 0.001 & 50.730 \\[-1em]
10 & 500 & 100 & 2 & 99.9 & 0.000 & 0.284 & 0 & 70 & 15.933 & 92.2 & 0.127 & 23.333 \\[-1em]
10 & 5000 & 100 & 2 & 100 & 0 & 0.614 & - & - & \textgreater24h & - & - & \textgreater24h \\[-1em]
10 & 50000 & 100 & 2 & 100 & 0 & 3.902 & - & - & \textgreater24h & - & - & \textgreater24h\\[-1em]
10 & 100 & 500 & 6 & 0.064 & 4.434 & 0.290 & 0 & 10.008 & 3.871 & 23.4 & 1.693 & 24.214 \\[-1em]
10 & 100 & 5000 & 6 & 0 & 6.684 & 0.646 & - & - & \textgreater24h & - & - &  \textgreater24h \\[-1em]
10 & 100 & 50000 & 6 & 0 & 8.553 & 3.728 & - & - & \textgreater24h & - & - & \textgreater24h\\[-1em]
10 & 100 & 500 & 10 & 85.9 & 0.250 & 0.288 & 0 & 10.971 & 3.914 & 69.2 & 0.689 & 15.457 \\[-1em]
10 & 100 & 5000 & 10 & 0 & 6.314 & 0.699 & - & - & \textgreater24h & - & - & \textgreater24h\\[-1em]
10 & 100 & 50000 & 100 & 91.1 & 0.207 & 3.767 & - & - & \textgreater24h & - & - & \textgreater24h\\
\hline\hline
\end{tabular}
\end{table}
\clearpage

\section{\label{sec:4} Case Study: Dimension Estimation in scRNA-seq Analysis}
The lung scRNA-seq data can be found on Gene Expression Omnibus (GEO) under accession code GSE136831. The data was examined by \citet{Adams759902} to build a single-cell atlas of Idiopathic Pulmonary Fibrosis (IPF). The data contains the gene expression of 312,928 cells profiled from 32 IPF lung samples, 18 chronic obstructive pulmonary disease (COPD) lung samples, and 29 control donor lung samples \citep{Adams759902}. We applied the dimension method to a subgroup of control lung sample (Patient ID 001C) with 2,000 highly variable genes and 127 cells. Figure 2 shows that among all the candidates with maximum double posterior probabilities (see Figure 2b), the eighth dimension is the largest with the maximum posterior probability of change point (see Figure 2a).

\begin{figure}[ht]
\includegraphics[width=\textwidth]{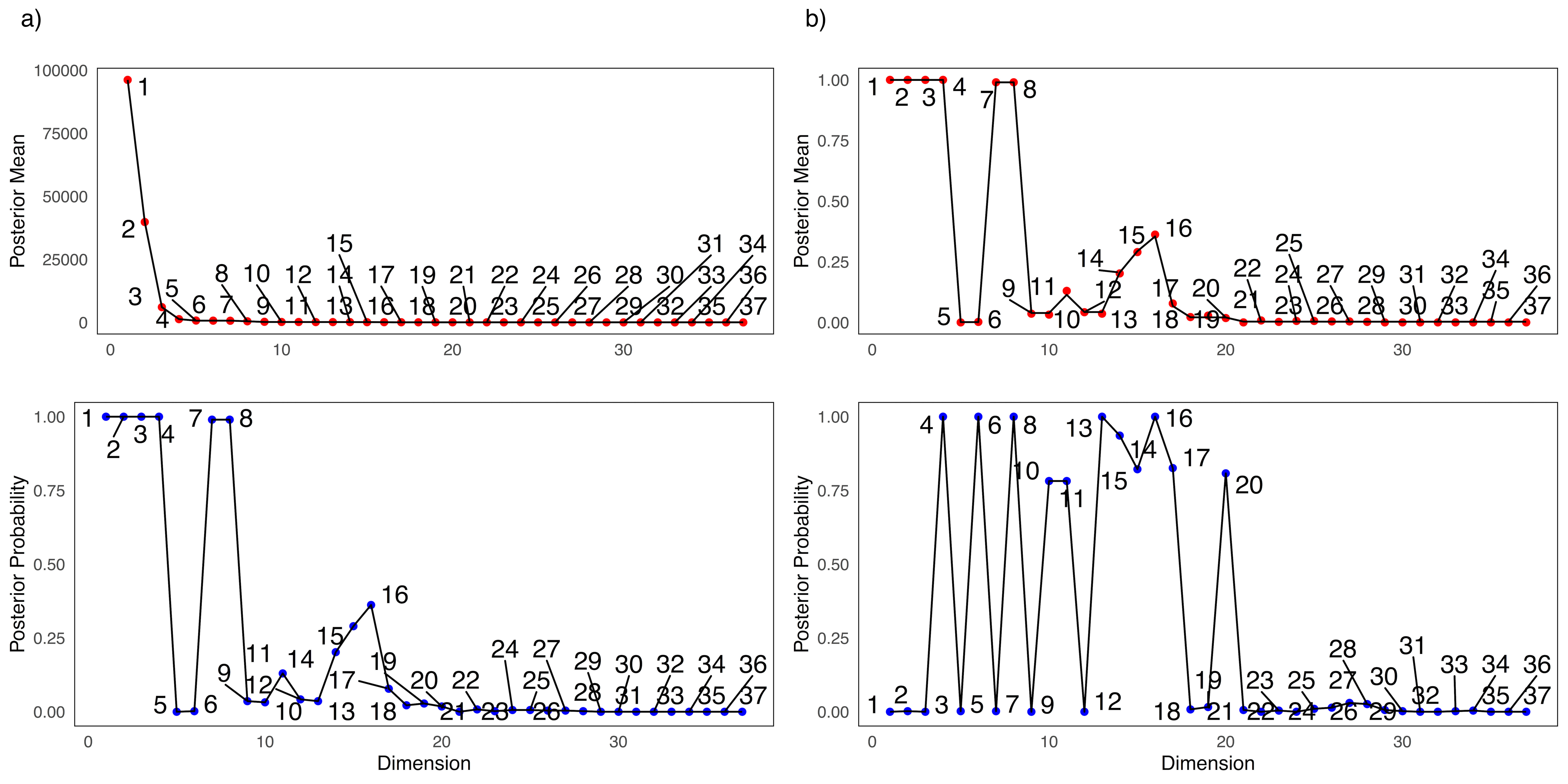}
\caption{\label{fig:FIG2}{(a) The posterior means of the deviation and posterior probability plots. (b) The posterior probability and double posterior probability plots.}}
\end{figure}

We then applied the Louvain clustering method implemented in the R package Seurat for PCs from Dimensions 1-50, 9-50, and 1-8 (see Figures 3a, 3b, 3c). The Uniform Manifold Approximation and Projection (UMAP)  for each plot confirms that major features of cell-type clustering are associated with Dimensions 1-8 and there is no new cluster formed from Dimensions 9-50. Clustering from the identified signal subspace identified two additional clusters that are not identified by Dimensions 1-50. To validate the clustering results, we identified marker genes in each cluster via differential expression analysis and further matched the clustering results to known cell types with identified marker genes (see Figure 3c for cell-type identification). Comparing to the original identification published by \citet{Adams759902} (see Figure 3d), we found that most of the cells types were correctly annotated with the small subset using Dimensions 1-8, with the exception of the T cells that were clustered with the NK cells. However, NK cells and T cells are phenotypically similar and they both derive from lymphoid lineage cells. By contrast, when we included noises from Dimensions 1-50, T cells, fibroblast, and secretory cells were misclassified as one cell type. Therefore, an accurate estimation of the signal dimension is paramount for cell-type identification to avoid clustering with an overabundance of noise or signal omission, both of which lead to misleading conclusions.

\begin{figure}[ht]
\includegraphics[width=\textwidth]{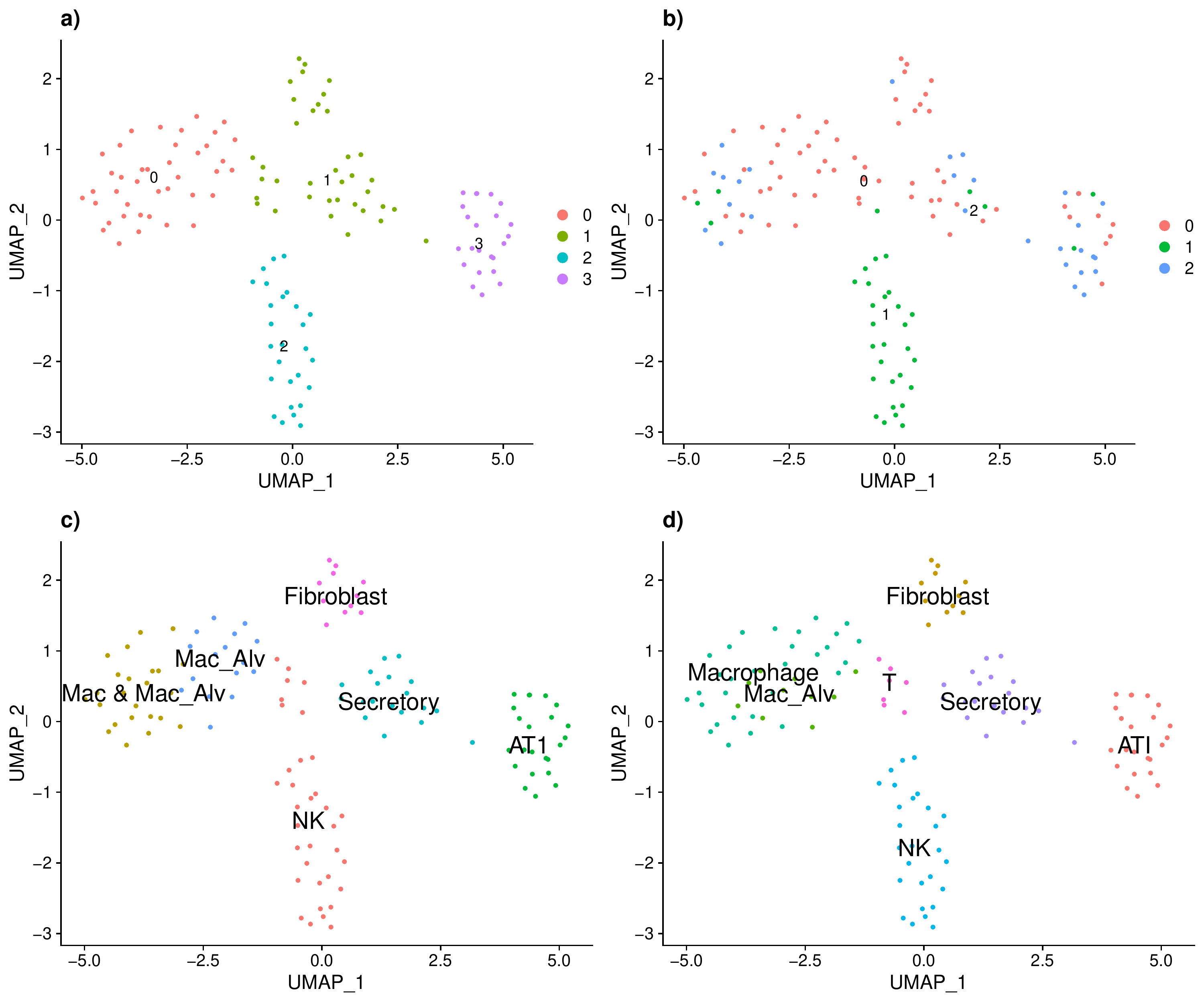}
\caption{\label{fig:FIG3} (a) UMAP visualization and cells labeled by Louvain clustering results with dimensions 1-50, (b) 9-50, and (c) 1-8. (d) Cell-type identification marked by \citet{Adams759902}.}
\end{figure}
\clearpage

\section{\label{sec:5} Conclusion}
In this chapter, we posed the problem of dimension estimation in a signal-plus-noise regime as the eigenvalue index where the difference between a data matrix and its inherent noise, defined by a random matrix, were close to zero. Based on this regime, we proposed a procedure to estimate the dimension of detectable signal, along with a derivation of distributional bounds for the magnitude of detectable signal required to be distinguished from noise. We implemented the proposed procedure and related methods as per the open-source R package at \url{https://github.com/WenlanzZ/dimension}. In simulation experiments, we demonstrated that the accuracy of the dimension method and its speed and computationally efficiency advantages over ladle and Hoff's method, both of which are essential for large and high-dimensional data sets. In high-dimensional cases with extremely low signal, the dimension method provided a robust low-rank estimation while the other two methods provided either zero rank or full rank. Our results showed that the dimension method is more accurate in estimating the dimension while remaining computationally scalable compared to current methods in a wide range of scenarios under both high- and low-dimensional settings. Besides making performance comparisons with the simulated data, we also demonstrated that community detection in the identified signal subspace was beneficial in cell-type identification when analyzing scRNA-seq data sets. In addition to the applications to genomic data sets, the dimension-estimation method has more general applications, such as digital imaging, speech modelling, and neural networks.

\begin{acknowledgments}

\end{acknowledgments}
\clearpage

\appendix
\section{Appendix: Proof of Equation 3}
\label{appendix:a}
Here we provide detailed proofs of the Equation \ref{dimsimp}. Base on the central limit theorem, we show that $\frac{1}{p}SN^T$ and $\frac{1}{p}NS^T \approx 0$. Given that $SN^T = NS^T$, we will focus on the proof for $\frac{1}{p}SN^T \approx 0$. The main technical challenge here comes from the case when the eigenvectors of $S$ and $N$ are not well-aligned ($U_N^T U_S \not\approx I$). Assume an extreme worst case when $S$ is a diagonal matrix with finite singular values $(\Sigma_s)_i$ for $i = 1, ...., p$ ($p - n$ rows are padded with zeros), we show that in general $\frac{1}{p}SN^T \not\approx 0$ for finite $p$. Let $N_{ij}$ be i.i.d. entries of matrix $N$ with $\E(N_{ij}) = 0$ and $\E({N^2}_{ij}) = \sigma^2$. Without loss of generality, we normalize $N_{ij}$ so that it has mean zero and variance one. 

\begin{equation}
\begin{array}{l}
\frac{1}{p}SN^T
=\frac{1}{p}\begin{bmatrix}
  S_{11}&0&\cdots &0 \\
  0&S_{22}&\cdots &0 \\
  \vdots & \vdots & \ddots & \vdots\\
  0&0&\cdots &S_{pp}
\end{bmatrix}\begin{bmatrix}
  N_{11}&N_{12}&\cdots &N_{1p} \\
  N_{21}&S_{22}&\cdots &N_{2p} \\
  \vdots & \vdots & \ddots & \vdots\\
  N_{p1}&0&\cdots &N_{pp}
  \end{bmatrix} \\
= \frac{1}{p}\sum_{i = 1}^{p} S_{ii} N_{ii}\\
= \frac{1}{p}\sum_{i = 1}^{p} \left( \Sigma_s \right )_i N_{ii} \\
\leq \frac{\Tr S}{p}\,.
\end{array}
\end{equation}
\noindent Thus we have $\frac{1}{p}SN^T \approx 0$ when $p\to\infty$ with a convergence rate $\frac{\Tr S}{p}$.
\clearpage

\section{Appendix: Derive moments for $W$}
\label{appendix:b}
Without loss of generality, we standardize the columns of random matrix $N_{n \times p}$ ($p > n$) by subtracting off the column-wise means and standardizing such that $\sigma^2 = 1$. We can first derive the first moment of $W$ as follows:
\begin{equation}
\begin{array}{l}
NN^T =\sum_{j = 1}^p N_{.j}N_{.j}^T
=\begin{bmatrix}
  \sum_{j = 1}^p N_{1j}^2 & \sum_{j = 1}^p N_{1j}N_{2j} & \cdots  & \sum_{j = 1}^p N_{1j}N_{nj} \\
  \sum_{j = 1}^p N_{2j}N_{1j} & \sum_{j = 1}^p N_{2j}^2 & \cdots & \sum_{j = 1}^p N_{2j}N_{nj} \\
  \vdots & \vdots & \ddots & \vdots\\
  \sum_{j = 1}^p N_{nj}N_{1j} & \sum_{j = 1}^p N_{nj}N_{2j} & \cdots & \sum_{j = 1}^p N_{nj}^2
\end{bmatrix}
\end{array}
\end{equation}
where the main diagonals 
\begin{equation}
\E \sum_{j = 1}^{p} N_{.j}^2 =  p\,,
\end{equation}
for $j = 1,2,...,p$. 
and the off-diagonal we have
\begin{equation}
\sum_{j = 1}^{p} \E N_{ij}N_{kj} = \sum_{j = 1}^{p} \E N_{ij} \E N_{kj} =  0 \,,
\end{equation}
for $j = 1,2,...,p$ and $i \neq k$.

\noindent Thus we have the first moment of the $W_p$ matrix as
\begin{equation}
\E W_p = \E \frac{1}{p}NN^T = I\,.
\end{equation}

\noindent We then derive the second moment of the $W_p$ matrix as follows:
\begin{equation}
\begin{array}{l}
NN^T NN^T = \\
\begin{bmatrix}
  \sum\limits_{j=1}^{p} \sum\limits_{i=1}^{n} \sum\limits_{k=1}^{p} N_{1k}N_{ik}N_{ij}N_{1j}
  & \sum\limits_{j=1}^{p} \sum\limits_{i=1}^{n} \sum\limits_{k=1}^{p} N_{1k}N_{ik}N_{ij}N_{2j}
  &\cdots 
  &\sum\limits_{j=1}^{p} \sum\limits_{i=1}^{n} \sum\limits_{k=1}^{p} N_{1k}N_{ik}N_{ij}N_{nj} \\
  \sum\limits_{j=1}^{p} \sum\limits_{i=1}^{n} \sum\limits_{k=1}^{p} N_{2k}N_{ik}N_{ij}N_{1j}
  &\sum\limits_{j=1}^{p} \sum\limits_{i=1}^{n} \sum\limits_{k=1}^{p} N_{2k}N_{ik}N_{ij}N_{2j}
  &\cdots 
  &\sum\limits_{j=1}^{p} \sum\limits_{i=1}^{n} \sum\limits_{k=1}^{p} N_{2k}N_{ik}N_{ij}N_{nj} \\
  \vdots & \vdots & \ddots & \vdots\\
  \sum\limits_{j=1}^{p} \sum\limits_{i=1}^{n} \sum\limits_{k=1}^{p} N_{nk}N_{ik}N_{ij}N_{1j}
  &\sum\limits_{j=1}^{p} \sum\limits_{i=1}^{n} \sum\limits_{k=1}^{p} N_{nk}N_{ik}N_{ij}N_{2j}
  &\cdots 
  &\sum\limits_{j=1}^{p} \sum\limits_{i=1}^{n} \sum\limits_{k=1}^{p} N_{nk}N_{ik}N_{ij}N_{nj}
  \end{bmatrix}
\end{array}
\end{equation}

\noindent where the expectation of $d_i^{th}$ main diagonal elements is
\begin{equation}
\begin{array}{l}
\E \sum\limits_{j=1}^{p} \sum\limits_{i=1}^{n} \sum\limits_{k=1}^{p} N_{d_ik}N_{ik}N_{ij}N_{d_ij} \\[10pt]
=p \E N_{d_ij}^4 +\sum\limits_{i=d_i,j \neq k} N_{d_ik}N_{d_ik}N_{d_ij}N_{d_ij}+ \sum\limits_{i \neq d_i, j=k} N_{d_ik}N_{ik}N_{ik}N_{d_ik} \\[10pt]
=p \E N_{d_ij}^4 +\sum\limits_{i=d_i,j \neq k}1+ \sum\limits_{i \neq d_i, j=k}1 \\[10pt]
=p(p+n+1)\,.
\end{array}
\end{equation}
Note that $\E N_{id_i}^4$ is the second moment of the chi-square distribution and the off-diagonal elements are with expectation 
\begin{equation}
\E\sum\limits_{j=1}^{p} \sum\limits_{i=1}^{n} \sum\limits_{k=1}^{p} N_{d_ik}N_{ik}N_{ij}N_{d_jj} = 0\,,
\end{equation}
where $d_i \neq d_j$.

\noindent Thus we have the second moment of $W_p$ as
\begin{equation}
\begin{array}{l}
\E {W_p}^2 = \E \frac{1}{p^2} NN^T NN^T
=(1+ \frac{1}{p}(n+1)) I\,.
\end{array}
\end{equation}
\clearpage

\section{Appendix: Proof of Theorem 1}
\label{appendix:c}
For a $N_{n \times p}$ matrix, where $p > n$. Let $N_{ij}$ be i.i.d. entries of $N$ with $\E(N_{ij}) = 0$, $\E({N^2}_{ij}) = \sigma^2$ and $\E({N^4}_{ij}) = \gamma^4 < \infty$. 

\noindent Consider the $j$th column of $N$, for $j = 1,2, ..., p$.  Let $\E N_{ij}^2 = \sigma^2$ and $Z_i = N_{ij}^2$, then we can write the main diagonal of the elements of the $W_p$ matrix as:
\begin{equation}
    \xi_j = \frac{1}{p} \left< N_{\cdot j}, N_{\cdot j} \right> = \frac{1}{p} \sum_{i=1}^p Z_i\,,
\end{equation}
where $\E Z_i = \sigma^2$ and $\Var \left( Z_i \right)= \gamma^4 - \sigma^4$. When $p$ sufficiently large to have the Lyapunov central limit effect, we have
\begin{equation}
 \xi_j \rightsquigarrow \mathcal{N} \left(\sigma^2, \frac{\gamma^4 - \sigma^4}{p} \right)\,.
\end{equation}
\clearpage

\section{Appendix: Proof of Theorem 2}
\label{appendix:d}
For a $N_{n \times p}$ matrix, where $p > n$. Let $N_{ij}$ be i.i.d. entries of $N$ with $\E(N_{ij}) = 0$, $\E({N^2}_{ij}) = \sigma^2$ and $\E({N^4}_{ij}) = \gamma^4 < \infty$.

\noindent Consider the $i$th row and $k$th column of $N$, for $1 \leq {i,k} \leq p$ and $i<k$. We can write the off-diagonal of the elements of the $W_p$ matrix at the $i$th row and $k$th column as
\begin{equation}
    \varepsilon_{ik} = \frac{1}{p} \sum_{j=1}^p N_{ij}N_{kj}\,.
\end{equation}
The product of two independent random variables can be written as
\begin{equation}
    N_{ij}N_{kj} = \frac{1}{4} (N_{ij}+N_{kj})^2 - \frac{1}{4} (N_{ij}-N_{kj})^2\,.
\end{equation}
We rewrite $N_{ij}N_{kj}$ as 
\begin{equation}
    N_{ij}N_{kj} = \frac{1}{4} (a^2- b^2)\,,
\end{equation}
where $a = N_{ij}+N_{kj}$ and  $b = N_{ij}-N_{kj}$.
The moments of $a$ and $b$ is derived as follows:
\begin{equation}
    \E a =  \E b = \sum_{j = 1}^n \mu_{j} = 0\,,
\end{equation}
\begin{equation}
    \E a^2 =  \E b^2 = \sum_{j = 1}^n \mu_{2j} = 2\sigma^2\,,
\end{equation}
\begin{equation}
    \E a^4 =  \E b^4 = \sum_{j = 1}^n \mu_{4j} + 6\sum_{j = 1}^{n-1}\sum_{k >j}^{n} \sigma_j^2\sigma_k^2 = 2\gamma^4 +6\sigma^4\,,
\end{equation}
\begin{equation}
    \Var a^4 =  \Var b^4 = 2\gamma^4 +2\sigma^4\,.
\end{equation}
In the next step, we show that $a$ and $b$ are independent. Consider a vector $U = \begin{bmatrix} N_{ij} \\ N_{kj} \end{bmatrix}$, $A = \begin{bmatrix} 1 & 1 \end{bmatrix}$ and $B = \begin{bmatrix} 1 & -1 \end{bmatrix}$. According to a theorem in multivariate probability that two linear functions $AU$ and $BU$ of a Gaussian vector $U$ are independent if an only if $A \Var(U)B^T = 0$.
\begin{equation}
    A \Var(U)B^T = \begin{bmatrix} 1 & 1 \end{bmatrix} \begin{bmatrix} \Var(N_{ij}) & \Cov(N_{ij}, N_{kj}) \\ 
     \Cov(N_{ij}, N_{kj}) & \Var(N_{kj}) \end{bmatrix} 
     \begin{bmatrix} 1 \\ -1 \end{bmatrix} = \Var(N_{ij}) - \Var(N_{kj}) = 0\,.
\end{equation}
\noindent Thus, $a$ and $b$ are independent. So that we can calculate the expectation of $N_{ij}N_{kj}$ as
\begin{equation}
    \E N_{ij}N_{kj} = \frac{1}{4} \E \left( a^2 - b^2 \right) = 0\,,
\end{equation}
and the variance as
\begin{equation}
    \Var N_{ij}N_{kj} = \frac{1}{16} \Var \left( a^2 - b^2 \right) = \frac{\gamma^4 + \sigma^4}{4}\,.
\end{equation}
Thus, for the off-diagonal entry $\varepsilon_{ik}$, the expectation is
\begin{equation}
    \E \varepsilon_{ik} = \frac{1}{p} \E \sum_{j=1}^p N_{ij}N_{kj} = 0\,,
\end{equation}
and the variance as
\begin{equation}
    \Var \varepsilon_{ik} =  \frac{1}{p^2} \Var\sum_{j=1}^p N_{ij}N_{kj} =\frac{\gamma^4 + \sigma^4}{4p}\,.
\end{equation}
When $p$ sufficiently large to have the Lyapunov central limit effect, we have
\begin{equation}
 \varepsilon_{ik} \rightsquigarrow \mathcal{N} \left(0, \frac{\gamma^4 + \sigma^4}{4p} \right)\,.
\end{equation}
\clearpage

\section{Appendix: Proof of Theorem 3}
\label{appendix:e}
Theorem \ref{theorem: diagnal distribution} indicates that the main diagonal entries $\xi_j$ $(j = 1, ..., n)$ of $W_p$ are iid. normally distributed random variables with mean $\sigma^2$ and variance $\frac{\gamma^4 - \sigma^4}{p}$. The Fisher–Tippett–\\Gnedenko theorem shows that for the maximum diagonal element 
\begin{equation}
\xi_{(n)} = max(\xi_j) \sim \mathbb{GEV}(\sigma^2,\frac{\sqrt{p(\gamma^4 - \sigma^4)}}{p}, 0)\,.
\end{equation}
The cumulative distribution function of the Generalized Extreme Value (GEV) distribution is
\begin{equation}
\Phi(\alpha_1) = \E \{ \xi_{(n)} > \alpha_1 \} = e^{-e^{-a}}\,,
\end{equation}
where $a = {\sqrt{p}(\alpha_1 - \sigma^2)}/{\sqrt{\gamma^4- \sigma^4}}$.

\noindent We can bound the the maximum diagonal element $\xi_{(n)}$ with the inequality $e^{x} \leq e^{x^2} + x$ for $x \in \mathbb{R}$ such that
\begin{equation}
    e^{-e^{-a}} \leq e^{e^{-2a}} -e^{-a}\,.
\end{equation}
Thus, we have
\begin{equation}
    \E \{ \xi_{(n)} > \alpha_1 \} \le e^{e^{-\frac{2\sqrt{p}(\alpha_1 - \sigma^2)}{\sqrt{\gamma^4- \sigma^4}}}} - e^{-\frac{\sqrt{p}(\alpha_1 - \sigma^2)}{\sqrt{\gamma^4- \sigma^4}}}\,.
\end{equation}
\clearpage

\section{Appendix: Proof of Theorem 4}
\label{appendix:f}
Theorem \ref{theorem: off-diagnal distribution} indicates that the off-diagonal entries $\varepsilon_{ik}$ $(i,k = 1, ..., n)$ and $i<k$ are iid. normally distributed random variables with mean $0$ and variance $\frac{\gamma^4 + \sigma^4}{4p}$. Given the distribution of the $i$th row and $k$th column of the off-diagonal element of the $W_p$ matrix. We can calculate the sum of the absolute values of the off-diagonal elements of the $k$th column of the $W_p$ matrix as $R_k = \sum_{i \neq k}|\varepsilon_{ik}|$. The absolute value of $\varepsilon_{ik}$ follows a half-normal distribution and the off-diagonal elements are independent, thus we can derive the expectation of the sum of the absolute off-diagonal values as follows:
\begin{equation}
 \mu_{GEV} = \E R_k = \sum_{i \neq k} \E |\varepsilon_{ik}| = (n-1) \sqrt{\frac{2(\gamma^4 + \sigma^4)}{4p\pi}} = \sqrt{\frac{(n-1)^2(\gamma^4 + \sigma^4)}{2p\pi}}\,,
\end{equation}
and the variance as
\begin{equation}
 \sigma^2_{GEV} = \Var R_k = \sum_{i \neq k} \Var |\varepsilon_{ik}| =  \frac{(n-1)(\pi - 2)(\gamma^4 + \sigma^4)}{4p\pi}\,.
\end{equation}
The Fisher–Tippett–Gnedenko theorem shows that for the maximum entry of $R_k$
\begin{equation}
R_{(n)} = max(R_k) \sim \mathbb{GEV}(\mu_{GEV}, \sigma_{GEV}, 0)\,,
\end{equation}
The cumulative distribution function of the Generalized Extreme Value (GEV) distribution is
\begin{equation}
\Phi(\alpha_2) = \E \{ R_{(n)} > \alpha_2 \} = e^{-e^{-b}}\,,
\end{equation}
where $b = \frac{\alpha_2 - \mu_{GEV}}{\sigma_{GEV}}$.
We can bound the $R_{(n)}$  with the inequality $e^{x} \leq e^{x^2} + x$ for $x \in \mathbb{R}$ such that
\begin{equation}
    e^{-e^{-b}} \leq e^{e^{-2b}} -e^{-b}\,.
\end{equation}
Furthermore, we can bound the eigenvalue of the $k$th column of the $W_p$ matrix by the Gershgorin circle theorem. Known that the eigenvalue of the $W_p$ matrix lies within at least one of the Gershgorin discs $D(\xi_k,R_k)$, where $D(\xi_k,R_k) \subseteq \mathbb{C}$ is a closed disc centered at the $k$th diagonal element of $W_p$ (denote as $\xi_k$) with radius $R_k$. With the distributional upper bound of the $\xi_k$ and distributional upper found of $R_k$, we have the upper bound of the largest eigenvalue of $W_p$, which lies with the Gershgorin discs $D(\xi_{(n)},R_{(n)})$ that are centered at $\xi_{(n)}$ with a radius $R_{(n)}$ of rate of convergence $\sqrt{\frac{n-1}{p}}$ for $p > n$.
\clearpage
\bibliography{references}

\begin{thebibliography}{20}
\def\enquote#1{``#1,''}
\def\plainquote#1{``#1''}
\expandafter\ifx\csname natexlab\endcsname\relax\def\natexlab#1{#1}\fi
\providecommand{\dourl}[1]{\href{http://#1}{\nolinkurl{#1}}}
\providecommand{\bibinfo}[2]{#2}
\providecommand{\noopsort}[1]{}
\providecommand{\switchargs}[2]{#2#1}
  \def\eatspace #1{#1}

\bibitem[{Adams \emph{et~al.}(2019)Adams, Schupp, Poli, Ayaub, Neumark,
  Ahangari, Chu, Raby, DeIuliis, Januszyk, Duan, Arnett, Siddiqui, Washko,
  Homer, Yan, Rosas, and Kaminski}]{Adams759902}
\bibinfo{author}{Adams, T.~S.}, \bibinfo{author}{Schupp, J.~C.},
  \bibinfo{author}{Poli, S.}, \bibinfo{author}{Ayaub, E.~A.},
  \bibinfo{author}{Neumark, N.}, \bibinfo{author}{Ahangari, F.},
  \bibinfo{author}{Chu, S.~G.}, \bibinfo{author}{Raby, B.},
  \bibinfo{author}{DeIuliis, G.}, \bibinfo{author}{Januszyk, M.},
  \bibinfo{author}{Duan, Q.}, \bibinfo{author}{Arnett, H.~A.},
  \bibinfo{author}{Siddiqui, A.}, \bibinfo{author}{Washko, G.~R.},
  \bibinfo{author}{Homer, R.}, \bibinfo{author}{Yan, X.},
  \bibinfo{author}{Rosas, I.~O.},  and \bibinfo{author}{Kaminski, N.}
  (\textbf{\bibinfo{year}{2019}}). \enquote{\bibinfo{title}{Single cell rna-seq
  reveals ectopic and aberrant lung resident cell populations in idiopathic
  pulmonary fibrosis}} \bibinfo{journal}{bioRxiv}
  \dourl{https://www.biorxiv.org/content/early/2019/09/06/759902},
  \dodoi{10.1101/759902}.

\bibitem[{Aparicio \emph{et~al.}(2018)Aparicio, Bordyuh, Blumberg, and
  Rabadan}]{aparicio2018quasi}
\bibinfo{author}{Aparicio, L.}, \bibinfo{author}{Bordyuh, M.},
  \bibinfo{author}{Blumberg, A.~J.},  and \bibinfo{author}{Rabadan, R.}
  (\textbf{\bibinfo{year}{2018}}). \enquote{\bibinfo{title}{Quasi-universality
  in single-cell sequencing data}} \bibinfo{journal}{arXiv preprint
  arXiv:1810.03602} .

\bibitem[{Barry and Hartigan(1993)}]{barry1993bayesian}
\bibinfo{author}{Barry, D.},  and \bibinfo{author}{Hartigan, J.~A.}
  (\textbf{\bibinfo{year}{1993}}). \enquote{\bibinfo{title}{A bayesian analysis
  for change point problems}} \bibinfo{journal}{Journal of the American
  Statistical Association} \textbf{88}(421), \bibinfo{pages}{309--319}.

\bibitem[{Carreira-Perpin{\'a}n(1997)}]{carreira1997review}
\bibinfo{author}{Carreira-Perpin{\'a}n, M.~A.} (\textbf{\bibinfo{year}{1997}}).
  \enquote{\bibinfo{title}{A review of dimension reduction techniques}}
  \bibinfo{journal}{Department of Computer Science. University of Sheffield.
  Tech. Rep. CS-96-09} \textbf{9}, \bibinfo{pages}{1--69}.

\bibitem[{Chung and Storey(2015)}]{chung2015statistical}
\bibinfo{author}{Chung, N.~C.},  and \bibinfo{author}{Storey, J.~D.}
  (\textbf{\bibinfo{year}{2015}}). \enquote{\bibinfo{title}{Statistical
  significance of variables driving systematic variation in high-dimensional
  data}} \bibinfo{journal}{Bioinformatics} \textbf{31}(4),
  \bibinfo{pages}{545--554}.

\bibitem[{Dumitriu and Edelman(2005)}]{dumitriu2005eigenvalues}
\bibinfo{author}{Dumitriu, I.},  and \bibinfo{author}{Edelman, A.}
  (\textbf{\bibinfo{year}{2005}}). \enquote{\bibinfo{title}{Eigenvalues of
  hermite and laguerre ensembles: large beta asymptotics}} in
  \emph{\bibinfo{booktitle}{Annales de l'IHP Probabilit{\'e}s et
  statistiques}}, Vol. 41, pp. \bibinfo{pages}{1083--1099}.

\bibitem[{Erdman \emph{et~al.}(2007)Erdman, Emerson
  \emph{et~al.}}]{erdman2007bcp}
\bibinfo{author}{Erdman, C.}, \bibinfo{author}{Emerson, J.~W.} \emph{et~al.}
  (\textbf{\bibinfo{year}{2007}}). \enquote{\bibinfo{title}{bcp: an r package
  for performing a bayesian analysis of change point problems}}
  \bibinfo{journal}{Journal of Statistical Software} \textbf{23}(3),
  \bibinfo{pages}{1--13}.

\bibitem[{Forrester and Warnaar(2008)}]{forrester2008importance}
\bibinfo{author}{Forrester, P.},  and \bibinfo{author}{Warnaar, S.}
  (\textbf{\bibinfo{year}{2008}}). \enquote{\bibinfo{title}{The importance of
  the selberg integral}} \bibinfo{journal}{Bulletin of the American
  Mathematical Society} \textbf{45}(4), \bibinfo{pages}{489--534}.

\bibitem[{Fr{\'e}chet(1927)}]{frechet1927loi}
\bibinfo{author}{Fr{\'e}chet, M.} (\textbf{\bibinfo{year}{1927}}).
  \enquote{\bibinfo{title}{Sur la loi de probabilit{\'e} de l'{\'e}cart
  maximum}} \bibinfo{journal}{Ann. Soc. Math. Polon.} \textbf{6},
  \bibinfo{pages}{93--116}.

\bibitem[{Gerschgorin(1931)}]{gerschgorin1931bounding}
\bibinfo{author}{Gerschgorin, S.} (\textbf{\bibinfo{year}{1931}}).
  \enquote{\bibinfo{title}{On bounding the eigenvalues of a matrix}}
  \bibinfo{journal}{Izv. Akad. Nauk. SSSR Otd Mat. Estest} \textbf{1},
  \bibinfo{pages}{749--754}.

\bibitem[{G{\"o}tze \emph{et~al.}(2004)G{\"o}tze, Tikhomirov
  \emph{et~al.}}]{gotze2004rate}
\bibinfo{author}{G{\"o}tze, F.}, \bibinfo{author}{Tikhomirov, A.} \emph{et~al.}
  (\textbf{\bibinfo{year}{2004}}). \enquote{\bibinfo{title}{Rate of convergence
  in probability to the marchenko-pastur law}} \bibinfo{journal}{Bernoulli}
  \textbf{10}(3), \bibinfo{pages}{503--548}.

\bibitem[{Hoff(2007)}]{hoff2007model}
\bibinfo{author}{Hoff, P.~D.} (\textbf{\bibinfo{year}{2007}}).
  \enquote{\bibinfo{title}{Model averaging and dimension selection for the
  singular value decomposition}} \bibinfo{journal}{Journal of the American
  Statistical Association} \textbf{102}(478), \bibinfo{pages}{674--685}.

\bibitem[{Kumar(2019)}]{kumar2019recursion}
\bibinfo{author}{Kumar, S.} (\textbf{\bibinfo{year}{2019}}).
  \enquote{\bibinfo{title}{Recursion for the smallest eigenvalue density of
  $\beta$-wishart--laguerre ensemble}} \bibinfo{journal}{Journal of Statistical
  Physics} \textbf{175}(1), \bibinfo{pages}{126--149}.

\bibitem[{Livan \emph{et~al.}(2018)Livan, Novaes, and
  Vivo}]{livan2018introduction}
\bibinfo{author}{Livan, G.}, \bibinfo{author}{Novaes, M.},  and
  \bibinfo{author}{Vivo, P.} (\textbf{\bibinfo{year}{2018}}).
  \emph{\bibinfo{title}{Introduction to random matrices: theory and practice}}
  (\bibinfo{publisher}{Springer}).

\bibitem[{Luo and Li(2016)}]{ladle}
\bibinfo{author}{Luo, W.},  and \bibinfo{author}{Li, B.}
  (\textbf{\bibinfo{year}{2016}}). \enquote{\bibinfo{title}{{Combining
  eigenvalues and variation of eigenvectors for order determination}}}
  \bibinfo{journal}{Biometrika} \textbf{103}(4), \bibinfo{pages}{875--887},
  \dourl{https://doi.org/10.1093/biomet/asw051}, \dodoi{10.1093/biomet/asw051}.

\bibitem[{Mar{\v{c}}enko and Pastur(1967)}]{Mar_enko_1967}
\bibinfo{author}{Mar{\v{c}}enko, V.~A.},  and \bibinfo{author}{Pastur, L.~A.}
  (\textbf{\bibinfo{year}{1967}}). \enquote{\bibinfo{title}{{DISTRIBUTION} {OF}
  {EIGENVALUES} {FOR} {SOME} {SETS} {OF} {RANDOM} {MATRICES}}}
  \bibinfo{journal}{Mathematics of the {USSR}-Sbornik} \textbf{1}(4),
  \bibinfo{pages}{457--483},
  \dourl{https://doi.org/10.10702Fsm1967v001n04abeh001994}.

\bibitem[{Vivo(2008)}]{vivo2008wishart}
\bibinfo{author}{Vivo, P.} (\textbf{\bibinfo{year}{2008}}).
  \enquote{\bibinfo{title}{From wishart to jacobi ensembles: statistical
  properties and applications}} Ph.D. thesis, \bibinfo{school}{Brunel
  University, School of Information Systems, Computing and Mathematics}.

\bibitem[{White and Smyth(2005)}]{white2005spectral}
\bibinfo{author}{White, S.},  and \bibinfo{author}{Smyth, P.}
  (\textbf{\bibinfo{year}{2005}}). \enquote{\bibinfo{title}{A spectral
  clustering approach to finding communities in graphs}} in
  \emph{\bibinfo{booktitle}{Proceedings of the 2005 SIAM international
  conference on data mining}}, \bibinfo{organization}{SIAM}, pp.
  \bibinfo{pages}{274--285}.

\bibitem[{Yang(1996)}]{yang1996asymptotic}
\bibinfo{author}{Yang, B.} (\textbf{\bibinfo{year}{1996}}).
  \enquote{\bibinfo{title}{Asymptotic convergence analysis of the projection
  approximation subspace tracking algorithms}} \bibinfo{journal}{Signal
  processing} \textbf{50}(1-2), \bibinfo{pages}{123--136}.

\bibitem[{Yong \emph{et~al.}(2013)Yong, Pearce
  \emph{et~al.}}]{yong2013beginner}
\bibinfo{author}{Yong, A.~G.}, \bibinfo{author}{Pearce, S.} \emph{et~al.}
  (\textbf{\bibinfo{year}{2013}}). \enquote{\bibinfo{title}{A beginner’s
  guide to factor analysis: Focusing on exploratory factor analysis}}
  \bibinfo{journal}{Tutorials in quantitative methods for psychology}
  \textbf{9}(2), \bibinfo{pages}{79--94}.

\end{thebibliography}
\end{document}